\documentclass{iopart}
\usepackage{iopams,setstack,graphicx,bbm}
\newcommand{\ket}[1]{\left| #1 \right\rangle}

\begin{document}

\title{Efficient formation of ground state ultracold molecules via
STIRAP from the continuum at a Feshbach resonance}

\author{Elena Kuznetsova and Susanne F. Yelin}
\address{Department of Physics, University of Connecticut,
Storrs, CT 06269}
\address{ITAMP, Harvard-Smithsonian Center
for Astrophysics, Cambridge, MA 02138} 
\author{Marko Gacesa, Philippe Pellegrini, and Robin C\^ot\'e}
\address{Department of Physics, University of Connecticut,
Storrs, CT 06269}
\date{\today}

\begin{abstract}
We develop a complete theoretical description of photoassociative Stimulated Raman Adiabatic Passage (STIRAP) 
near a Feshbach resonance in a thermal atomic gas. We show that it is possible to use \textit{low intensity} laser pulses to 
directly excite the continuum at a Feshbach resonance and transfer nearly the entire atomic population to the lowest 
rovibrational level in the molecular ground state. In case of a broad resonance, commonly found in several diatomic alkali 
molecules, our model predicts a transfer efficiency $\eta$ up to 97\% for a given atom pair, and up to 70\% when averaged over 
an atomic ensemble. The laser intensities and pulse durations needed for optimal transfer are $10^2-10^3$ W/cm$^2$ and several $\mu$s. Such efficiency compares to or surpasses currently available techniques for creating stable diatomic molecules, and the versatility of this approach simplifies its potential use for many molecular species.

\end{abstract}

\maketitle

\section{Introduction}
The realization of rovibrationally stable dense samples of ultracold diatomic molecules remains one of the major goals 
in the field of atomic and molecular physics. While cooling diatomic alkali molecules was seen as a logical next step 
following the optical cooling of atoms, many of the possible applications currently under investigation extend beyond 
atomic and molecular physics. Testing fundamental symmetries based on high-precision spectroscopy of ultracold molecules 
\cite{DeMille-EDM,Hudson,weak-inter} or the attempts to detect the time variation of fundamental constants \cite{fine-structure} 
are examples of such applications. Another one is ultracold chemistry, where the interacting species and products are in a 
coherent quantum superposition state and could be realized by controlling reactive collisional processes \cite{quant-chem}. 
Important insights about new phases of matter could be gained from strong anisotropic dipole-dipole interaction between 
ultracold dipolar molecules \cite{Baranov}. Finally, ultracold polar molecules could also represent an attractive platform for 
quantum computation \cite{DeMille,Susanne}. Many of those applications require dense samples of ultracold polar molecules in the lowest rovibrational state that makes them collisionally stable and long-lived.

Translationally ultracold (100 nK - 1 mK) molecules are produced from an ultracold atomic gas by photoassociation (PA) \cite{photoass} or magnetoassociation (MA) \cite{Feshbach}. In a typical PA scheme, a pair of colliding atoms is photoassociated into a bound electronically excited molecular state that spontaneously decays, forming molecules in the electronic ground state. In magnetoassociation, a magnetic field is adiabatically swept across a Feshbach resonance, converting two atoms in a matching scattering state into a molecule. Both techniques produce weakly bound molecules in highly excited vibrational states of the ground electronic potential. Such molecules have to be rapidly transferred to deeply bound vibrational states before they are lost from the trap due to inelastic collisions.

Stimulated Raman Adiabatic Passage (STIRAP) \cite{STIRAP} has recently attracted significant interest as an efficient way 
to produce deeply bound molecules, starting from Feshbach molecules \cite{STIRAP-Fesh,STIRAP-Fesh1}. It allows to realize high transfer 
efficiency and preserve the high phase-space density of an initial atomic gas. In STIRAP, the laser pulses, coupling an 
initial and a final state to an intermediate excited state, are applied in a counter-intuitive sequence where a pump pulse is 
preceeded by a Stokes pulse. During the transfer, the system stays in a "dark" state, {\it i.e.}, a coherent superposition of 
initial and final states, preventing any losses that would otherwise occur from the excited state. By adiabatically changing amplitudes of the laser pulses, the "dark" state evolves from the initial to the final state, resulting in nearly 100\% transfer efficiency \cite{STIRAP}.

Efficient adiabatic passage from the continuum requires laser pulses shorter than the coherence time of the continuum 
\cite{Vardi,Vardi-OE,Ye-photoass}. The adiabaticity condition of STIRAP, $\Omega \tau_{\mathrm{tr}} \gg 1$, where $\tau_{\mathrm{tr}}$ is the transfer time, 
therefore implies a large effective Rabi frequency $\Omega$ for the pulses. In addition, dipole matrix elements between the continuum and the bound state are usually small, and so the pump pulse that couples the continuum and the excited state would require a very high intensity, which proves impractical. Thus the previous STIRAP experiments \cite{STIRAP-Fesh}, being restricted by the very short coherence time of the continuum, used a Feshbach molecular state as an initial state.

The small continuum-bound dipole matrix elements can be dramatically increased by photoassociating atoms in the vicinity of a 
Feshbach resonance. It has been shown, both theoretically and experimentally, that the photoassociation rate increases in the 
presence of a Feshbach resonance by several orders of magnitude \cite{PA-Fesh,PA-Fesh1,PA-Fesh2,FOPA}.
This can be explained by considering that delocalized scattering states acquire some bound-state character due to admixture of a 
bound level associated with a closed channel, resulting in a large increase of the Franck-Condon factor between the initial scattering state and the final excited state. The recently proposed Feshbach Optimized Photoassociation (FOPA) technique \cite{FOPA} relies on this enhancement to directly reach deeply bound ground state vibrational levels from the scattering continuum. Consequently, photoassociation in the vicinity of a Feshbach resonance
is expected to increase molecular formation rate up to $10^{6}$ molecules/s \cite{FOPA}.

In the present work, we combine the approach used in FOPA with STIRAP for reducing the required pulse intensity. 
We predict highly efficient transfer of an entire atomic ensemble into the lowest rovibrational level in the molecular 
ground state.

The paper is organized as follows. In Section II, we derive a theoretical model of a combined
atomic and molecular system. Fano theory is used to describe the interaction of a
bound molecular state with the scattering continuum, represented as closed and open channel, respectively. 
The resulting continuum states are coupled by two laser fields to
the vibrational target state in the ground state via the intermediate excited molecular electronic vibrational state. In Section III, we present the results of numerical solutions of the model for several alkali dimers. We find optimal Rabi frequencies and profiles of STIRAP pulses for those systems. Finally, we conclude in Section IV.

\section{Model}

\begin{figure}
\begin{center}
\includegraphics[width=0.7\textwidth,clip]{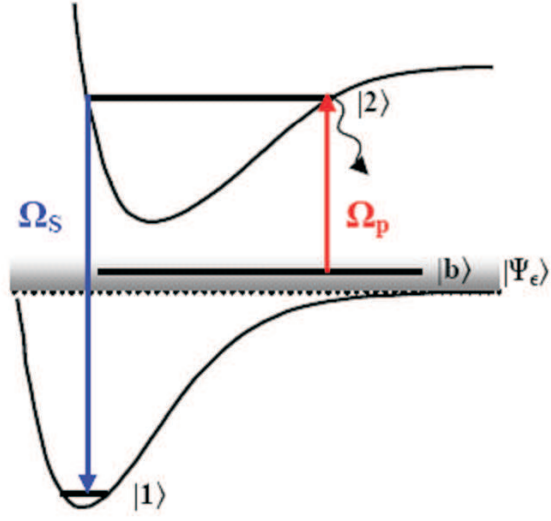}
\caption{Schematics: a population from the initial state $\ket{\Psi_{\epsilon}}$ is transferred to a final target state $\ket{1}$ via an intermediate state $\ket{2}$. Both $\ket{\Psi_{\epsilon}}$ and $\ket{1}$ are coupled to $\ket{2}$ by a pump and a Stokes pulse, respectively labeled $\Omega_P$ and $\Omega_S$. A bound level $\ket{b}$ corresponding to a closed channel can be imbedded in the continuum.}
\label{Levels}
\end{center}
\end{figure}

We consider a three level system as represented in Figure~\ref{Levels}. The ground level labeled $\ket{1}$ is the final 
product state to which a maximun of population must be transfered. Typically, this level will be the lowest virational 
level ($v''=0,J''=0$) of a ground molecular potential. This ground level is coupled to an excited bound level $\ket{2}$ of an 
excited molecular potential via a ''Stokes'' pulse depicted by the blue down-arrow in Figure~\ref{Levels}. This level 
$\ket{2}$ is itself coupled via a pump pulse (red up-arrow) to an initial continuum of unbound scattering states 
$\ket{\Psi_{\epsilon}}$ of energies $\epsilon$ (grey area in Figure~\ref{Levels}). If we denote $C_{1}$, $C_{2}$ and 
$C(\epsilon)$ the time dependent amplitudes associated to the final, intermediate, and initial states $\ket{1}$, $\ket{2}$, 
and $\ket{\Psi_{\epsilon}}$, respectively, then the total wave function $\ket{\Phi}$ of the system is given by:

\begin{equation}
 \ket{\Phi}=C_{1}\ket{1}+C_{2}\ket{2}+\int \; d\epsilon \; C(\epsilon)\ket{\Psi_{\epsilon}}.
\end{equation}

No restriction applies to the definition of the continuum state $\ket{\Psi_{\epsilon}}$ as it can be associated to 
either a single-channel or a multi-channel scattering state. 
In this work, we consider the multi-channel case in which a bound level $\ket{b}$ associated to a closed channel is 
embedded in the continuum of scattering states $\ket{\epsilon'}$ of an open channel. When the energy of $\ket{\epsilon'}$ 
coincides with that of $\ket{b}$, a so-called Feshbach resonance \cite{Fesh} occurs. These are common in binary collisions of 
alkali atoms due to hyperfine mixing and the tuning of the Zeeman interaction by an external magnetic field, hence the 
possibility to control interatomic interactions with a magnetic field.
Following the Fano theory presented in Ref. \cite{Fano}, the scattering state $\ket{\Psi_{\epsilon}}$ can be expressed as:

\begin{equation}
\ket{\Psi_{\epsilon}}=a(\epsilon)\ket{b}+\int d\epsilon' \; b(\epsilon,\epsilon')\ket{\epsilon'}\;,
\end{equation}
with 
\begin{equation}
 a(\epsilon)=\sqrt{\frac{2}{\pi\Gamma(\epsilon)}}\sin\Delta \;,
\end{equation}
and 
\begin{equation}
b(\epsilon,\epsilon')=\frac{1}{\pi}\sqrt{\frac{\Gamma(\epsilon')}{\Gamma(\epsilon)}}\frac{\sin\Delta}{\epsilon-\epsilon'}-\cos\Delta \; \delta(\epsilon-\epsilon') \;.
\end{equation}
Here, $\Delta =-\arctan(\frac{\Gamma}{2(\epsilon-\epsilon_{F})})$ is the phase shift due to the interaction between $\ket{b}$ and the scattering state $\ket{\epsilon}$ 
of the open channel. We assume $\Delta \in [-\pi/2,\pi/2]$. The width of the Feshbach resonance,
$\Gamma=2\pi|V(\epsilon)|^{2}$, is weakly dependent on the energy, while $V(\epsilon)$ is the interaction strength between the open and closed channels. The position of the resonance,
$\epsilon_{F}=E_{b}+P\int\frac{|V(\epsilon')|^{2}d\epsilon'}{\epsilon -\epsilon'}$, includes an interaction induced shift from the energy of the bound state $E_{b}$.

If we label $E_i$ the energy of the state $\ket{i}$, the total Hamiltonian $H$ is given by:
\begin{equation}
  H=\sum_{i=1,2} E_{i} |i\rangle \langle i| +
  \int d\epsilon \; \epsilon |\Psi_{\epsilon}\rangle \langle \Psi_{\epsilon}| 
  +V_{\mathrm{light}} \;.
\end{equation}
The light-matter interaction Hamiltonian $V_{\mathrm{light}}$ takes the form:
\begin{equation}
V_{\mathrm{light}}=-\vec{\mu}_{21}\cdot\vec{{\cal E}}_{S}
|2\rangle \langle 1|
-\int d\epsilon\; \vec{\mu}_{2\Psi_{\epsilon}}\cdot\vec{{\cal E}}_{p}\ket{2}
\langle\Psi_{\epsilon}|  + \mathrm{H.c.}\; ,
\end{equation}
where $\vec{{\cal E}}_{p,S}=\hat{\vec{e}}_{p,S}{\cal E}_{p,S}\exp(-i\omega_{p,S}t)+\mathrm{c.c.}$ 
are the pump and Stokes laser fields of polarization $\hat{\vec{e}}_{p,S}$, respectively, while $\vec{\mu}_{21}$ and $\vec{\mu}_{2\Psi_{\epsilon}}$ are the dipole transition moments between the states $\ket{2}$ and $\ket{1}$, and $\ket{2}$ and $\ket{\Psi_{\epsilon}}$, respectively.
In this form the Hamiltonian already takes into account mixing between the bound state of the closed channel and scattering states 
of the open channel. The Schr\"odinger equation describing STIRAP conversion of two atoms into a molecule is:
\begin{eqnarray}
i\hbar \frac{\partial C_{1}}{\partial t} & = & E_{1}\;C_{1}-\vec{\mu}^{*}_{21}\cdot\vec{{\cal E}}^{*}_{S}\;C_{2},\\
i\hbar \frac{\partial C_{2}}{\partial t} & = & E_{2}\;C_{2}-\vec{\mu}_{21}\cdot\vec{{\cal E}}_{S}\;C_{1}-\int_{\epsilon_{th}}^{\infty}d\epsilon\; \vec{\mu}_{2\Psi_{\epsilon}}\cdot\vec{{\cal E}}_{p}\;C(\epsilon),\\
i\hbar \frac{\partial C(\epsilon)}{\partial t} & = & \epsilon \;C(\epsilon)-\vec{\mu}^{*}_{2\Psi_{\epsilon}}\cdot\vec{{\cal E}}^{*}_{p}\;C_{2}.
\end{eqnarray}

For simplicity, we set the origin of the energy to be the position of the ground state $\ket{1}$, and use the rotating wave 
approximation with $C_1=c_1$, $C_2=c_2e^{-i\omega_S t}$, and $C(\epsilon)=c(\epsilon)e^{-i(\omega_S-\omega_P)t}$. The Schr\"odinger equation becomes:
\begin{eqnarray}
\label{eq:c1}
i \frac{\partial c_{1}}{\partial t} & = & -\Omega_{S}c_{2},\\
\label{eq:c2}
i\frac{\partial c_{2}}{\partial t} & = & \delta c_{2}-\Omega_{S}c_{1}-\int_{\epsilon_{th}}^{\infty} d\epsilon\; \Omega_{\epsilon}c(\epsilon),\\
\label{eq:cont-ampl}
i\frac{\partial c(\epsilon)}{\partial t} & = & \Delta_{\epsilon} c(\epsilon)-\Omega^{*}_{\epsilon}c_{2},
\end{eqnarray}
where $\delta=E_{2}/\hbar -\omega_{S}$, $\Delta_{\epsilon}=\epsilon/\hbar -(\omega_{S}-\omega_{p})$, and $\epsilon_{th}$ is the dissociation 
energy of the ground electronic potential with respect to the state $\ket{1}$. 
The Rabi frequencies of the fields are $\Omega_{S}=\vec{\mu}_{21}\cdot\hat{\vec{e}}_{S}{\cal E}_{S}/\hbar$ (assumed real), 
$\Omega_{\epsilon}=\vec{\mu}_{2\Psi_{\epsilon}}\cdot\hat{\vec{e}}_{p}{\cal E}_{p}/\hbar$.

The previous system of three equations can be reduced into a two-equation system by eliminating the continuum 
amplitude $c(\epsilon)$ in Eq.(\ref{eq:cont-ampl}). Introducing a solution in the form of $c(\epsilon)=s(\epsilon)\exp{(-i\Delta_{\epsilon}t)}$ into Eq.(\ref{eq:cont-ampl}), we get
\begin{equation}
   s=i\int^{t}_{0}dt'\;\Omega^{*}_{\epsilon}(t')c_{2}(t')e^{i\Delta_{\epsilon}t'}+s_{\epsilon}(t=0) ,
\end{equation}
where $t=0$ is some moment before the collision of the two atoms. The resulting continuum amplitude is 
\begin{equation}
c=i\int^{t}_{0}dt'\;\Omega^{*}_{\epsilon}(t')c_{2}(t')e^{i\Delta_{\epsilon}(t'-t)}+s_{\epsilon}(t=0)e^{-i\Delta_{\epsilon}t}.
\end{equation} 

Inserting this result into Eq. (\ref{eq:c2}), we obtain a final system of equations for the amplitudes of the bound states:
\begin{eqnarray}
\label{eq:c1-new}
i \frac{\partial c_{1}}{\partial t}  &=&  -\Omega_{S}c_{2} ,\\
\label{eq:c2-new}
i\frac{\partial c_{2}}{\partial t} 
&=& \delta c_{2}-\Omega_{S}c_{1}+i\int_{\epsilon_{th}}^{\infty}d\epsilon\;\Omega_{\epsilon}(t)\int^{t}_{0}dt'\;\Omega_{\epsilon}(t')^{*}c_{2}(t')e^{i\Delta_{\epsilon}(t'-t)} \nonumber \\ 
&&-\int_{\epsilon_{th}}^{\infty}d\epsilon\; \Omega_{\epsilon}(t)s_{\epsilon}(t=0)e^{-i\Delta_{\epsilon}t} , \\
&\equiv& \delta c_{2}-\Omega_{S}c_{1}+T-S .\nonumber
\end{eqnarray}\\

The third term of Eq. (\ref{eq:c2-new}), labelled $T$, corresponds to the back-stimulation term, whereas the last term, labelled $S$, corresponds to the source function. In this source term, 
the initial amplitude of the continuum wave function $s_{\epsilon}(t=0)$
describing a collision at $t_{0}$ of two atoms with relative energy $\epsilon_{0}$ has been discussed in various contributions \cite{Vardi,Vardi-OE,Ye-photoass}. A Gaussian wavepacket provides the most classical description of a two-atom collision 
characterized by a minimal uncertainty relation between the energy bandwidth 
$\delta_{\epsilon}$ of the wavepacket and the duration of the collision: 
\begin{equation}
\label{s_epsiton_t0}
s_{\epsilon}(t=0)=\frac{1}{(\pi \delta^{2}_{\epsilon})^{1/4}}e^{-\frac{(\epsilon -\epsilon_{0})^{2}}{2\delta^{2}_{\epsilon}}+\frac{i}{\hbar}(\epsilon -\epsilon_{0})t_{0}}.
\end{equation}

Futhermore, the Rabi frequency of the field coupling continuum states $\ket{\Psi_{\epsilon}}$ to the state $\ket{2}$ is given by \cite{Fano}
\begin{equation}
\label{eq:Rabi-fr}
\Omega_{\epsilon}=\frac{\vec{\mu}_{2\epsilon}\cdot\hat{\vec{e}}_{p}{\cal E}_{p}}{\hbar}\frac{q\Gamma/2+\epsilon -\epsilon_{F}}{\sqrt{(\Gamma/2)^{2}+(\epsilon - \epsilon_{F})^{2}}}\mathrm{sgn}(\epsilon-\epsilon_{F}),
\end{equation}
where $\vec{\mu}_{2\epsilon}$ is the dipole matrix element 
between an unperturbed scattering state $\ket{\epsilon}$ and the 
state $\ket{2}$, and $q$ is the Fano parameter, expressed as:
\begin{equation}
\label{eq:q}
q=\frac{(\vec{\mu}_{2b}\cdot \hat{\vec{e}}_{p})+P\int\frac{V(\epsilon')(\vec{\mu}_{2\epsilon'}\cdot \hat{\vec{e}}_{p})d\epsilon'}{\epsilon-\epsilon'}}{\pi V^{*}(\epsilon)(\vec{\mu}_{2\epsilon}\cdot \hat{\vec{e}}_{p})},
\end{equation}
where $\hat{\vec{e}}_{p}$ is the polarization vector of the pump field, and $\vec{\mu}_{2b}$ is the dipole matrix element between bound states $\ket{2}$ and $\ket{b}$. 
The $q$ factor is essentially the ratio of the dipole matrix elements from the state 
$\ket{2}$ to the bound state $\ket{b}$ (modified by the continuum) and to an unperturbed continuum
state $\ket{\epsilon}$. This factor can be made much larger than unity, and as will be shown below, the total dipole matrix element from the continuum can be enhanced by this factor in the presence of the resonance. The magnitude of $q$ can be controlled by the choice of the vibrational state $\ket{2}$. 
Selecting a tightly bound excited vibrational state will increase the bound-bound and decrease the continuum-bound dipole matrix elements, resulting in larger $q$. On the contrary, choosing a highly excited state close to a dissociation threshold decreases $q$.

Using the expressions given in Eqs.(\ref{s_epsiton_t0}), (\ref{eq:Rabi-fr}), and (\ref{eq:q}) for the initial amplitude 
of the continuum wave function, the Rabi frequency between the continuum state $\ket{\Psi_{\epsilon}}$ and the excited bound 
state $\ket{2}$, and the Fano parameter, respectively, we obtain the following complete expression for the source term:
\begin{equation}
\label{eq:S}
S=S_{0}\int_{\epsilon_{th}}^{\infty}d\epsilon \;g(q,\epsilon)\; \mathrm{sgn}(\epsilon-\epsilon_{F}) 
e^{-\frac{(\epsilon -\epsilon_{0})^{2}}{2\delta^{2}_{\epsilon}}+\frac{i(\epsilon -\epsilon_{0})t_{0}}{\hbar}}e^{-i\Delta_{\epsilon}t},
\end{equation} 
with $S_{0}=\vec{\mu}_{2\epsilon}\cdot\hat{\vec{e}}_{p}{\cal E}_{p}/\hbar(\pi \delta_{\epsilon})^{1/4}$, and where the function $g(q,\epsilon)$ is defined as
\begin{equation}
  g(q,\epsilon)\equiv \frac{q+\frac{2}{\Gamma}(\epsilon-\epsilon_{F})}{\sqrt{1+\frac{4}{\Gamma^2}(\epsilon-\epsilon_{F})^{2}}} .
 \label{eq:g}
\end{equation}
We assume that the unperturbed continuum is structureless and the coresponding Rabi frequency 
$\vec{\mu}_{2\epsilon}\cdot\hat{\vec{e}}_{p}{\cal E}_{p}/\hbar$ depends only weakly on the energy. We also extend 
$\epsilon_{th}$ to $-\infty$ to have the initial continuum wavefunction normalized to unity: 
$\int_{-\infty}^{\infty}d\epsilon \; |C(\epsilon)|^{2}=1$.

We can as well obtain a complete expression for the back-stimulation term $T$. We have:
\begin{equation}
 T=\left|\frac{\vec{\mu}_{2\epsilon}\hat{\vec{e}}_{p}}{\hbar}\right|^{2}{\cal E}_{p}(t)\int_{\epsilon_{th}}^{\infty}d\epsilon \; g^2(q,\epsilon)
 \int_{0}^{t} dt'\;c_{2}(t'){\cal E}_{p}(t')e^{i\Delta_{\epsilon}(t'-t)} .
\end{equation}
Extending the lower integration limit allows for an analytical solution for the integrals over energy and time, 
leading to the following expression for the back-stimulation term:
\begin{eqnarray}
T&=&\label{eq:back-stimulation}
\left|\frac{\vec{\mu}_{2\epsilon}\hat{\vec{e}}_{p}}{\hbar}\right|^{2}
\left [\pi \hbar {\cal E}_{p}^{2}(t)c_{2}(t)+\frac{\pi \Gamma}{2}(q-i)^{2}{\cal E}_{p}(t) \right. \nonumber\\ 
&&  \left.  \times\int_{0}^{t}dt'\; c_{2}(t')
{\cal E}_{p}(t')e^{[\Gamma/2\hbar+i(\epsilon_{F}/\hbar-\omega_{S}+\omega_{p})]
(t'-t)} \right] .
\end{eqnarray}

\section{Results}

In this work, we consider two different cases: first, when $\Gamma \gg \delta_{\epsilon}$, {\it i.e.}, when the width $\Gamma$ of 
the Feshbach resonance is much larger than the thermal energy spread $\delta_{\epsilon}$ of the colliding atoms, and second 
when $\Gamma \ll \delta_{\epsilon}$. By
considering these two limiting cases of broad and narrow resonances, more practical expressions for both the source term $S$ and the back stimulation term $T$ can be found. The derivation of the final system of equations is given in \ref{appendixA}. Here, we only describe the solutions of these systems for both broad and narrow resonances.

\begin{figure}[h]
\begin{center}
\includegraphics[width=\textwidth,clip]{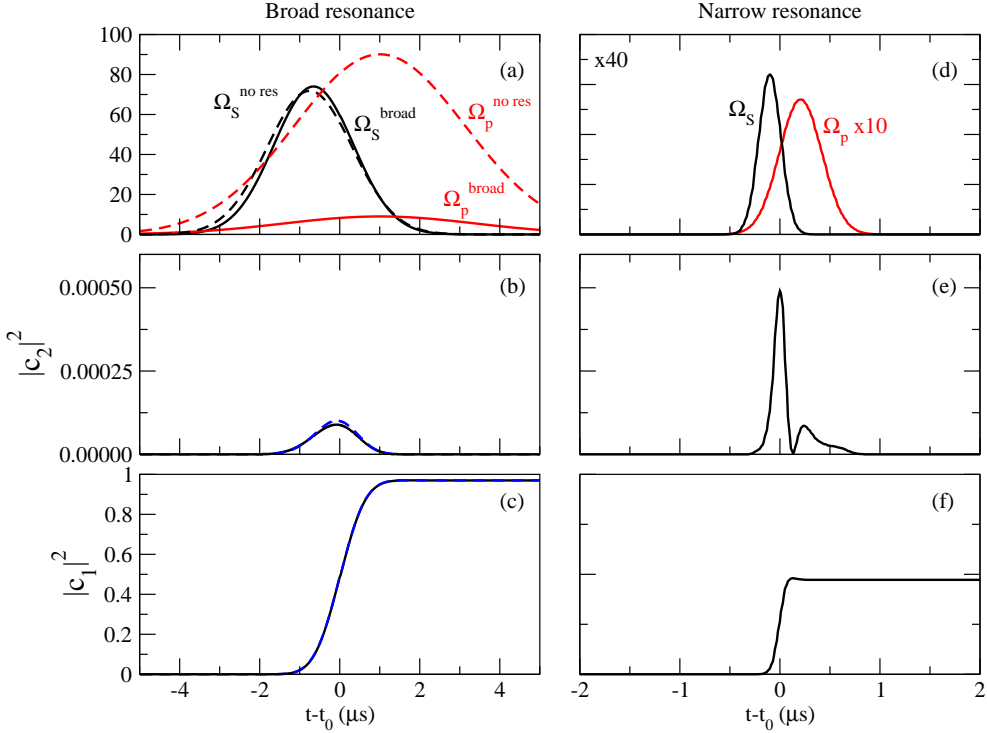}
\end{center}
\caption{Time-dependence of the Stokes and pump pulses (top row) and population in state $|2\rangle$ 
(middle row) and target state $|1\rangle$ (bottom row) for the STIRAP sequence. The left column is for a broad Feshbach 
resonance, while the right column is for a narrow resonance (see Table~\ref{table:res-par} for values of parameters used). 
The dashed blue lines in the left column are the results obtained without resonance, when the parameters are adjusted to obtain 
the same overall transfer fraction as for the broad resonance. Stokes Rabi frequency is in units of $10^{6}$ s$^{-1}$, while 
the pump Rabi frequency is in dimensionless units $\left(16\pi/\delta_{\epsilon}\right)^{1/4}\vec{\mu }_{2\epsilon}\vec{e}_{p}{\cal E}_{p}$ in 
the broad resonance limit and $\left(2\pi/\Gamma\right)^{1/2}\vec{\mu}_{2\epsilon}\vec{e}_{p}{\cal E}_{p}$ in the 
narrow resonance limit. Note that the scale for the Rabi frequencies in the narrow resonance case is 40 times the scale for the broad 
resonance, and the magnitude of the pump Rabi frequency is enlarged 10 times for better visibility.}
\label{resonances1}
\end{figure}

Using the parameters of the Stokes and pump photoassociating pulses listed in Table~\ref{table:res-par} for a broad 
($\Gamma = 1$ mK) and a narrow ($\Gamma = 1$ $\mu$K) Feshbach resonance, we obtain the results depicted in 
Fig.~\ref{resonances1}, with the left column corresponding to the broad resonance, and the right column to the narrow resonance. 
The top row shows the variation of the Rabi frequencies over the time period required for the population transfer calculated 
using Eqs.(\ref{eq:c1-new}) and (\ref{eq:c2-new}) along with population in the intermediate state $\ket{2}$ (middle row) and final state $\ket{1}$ (bottom row).

For the broad case, we considered a Feshbach resonance with a width $\Gamma=1$ mK, which is a typical value for broad resonances 
(see examples in Appendix A.1), 
and a thermal atomic ensemble with an energy bandwidth $\delta_{\epsilon}=10$ $\mu$K. 
We see that the transfer can reach $\sim 97$\% of the continuum state into the target state $|1\rangle$ 
(see Fig.~\ref{resonances1} c). The parameters of the Gaussian laser pulses we used (optimized Rabi frequencies, durations and delays of laser pulses) are 
given in Table~\ref{table:res-par}: the peak intensities of the Stokes and pump fields were 
calculated from Rabi frequencies as $I_{S}=c{\cal E}^{2}_{S}/8\pi=c(\Omega_{S}^{0}\hbar)^{2}/8\pi \mu^{2}_{21}$ and 
$I_{p}=c{\cal E}^{2}_{p}/8\pi=c(\Omega_{p}^{0})^{2}\delta_{\epsilon}/32\pi^{3/2}\mu_{2\epsilon}^{2}$, where we use 
Eq.(\ref{eq:q}) to estimate the continuum-bound dipole matrix element $\mu_{2\epsilon}\approx \mu_{2b}/q \pi V(\epsilon)=\sqrt{2}\mu_{2b}/q\sqrt{\pi \Gamma}$, resulting in $I_{p}=q^{2}c (\Omega_{p}^{0})^{2}\delta_{\varepsilon}\Gamma/64\sqrt{\pi}\mu^{2}_{2b}$. 

\begin{table}
\centering
\caption{Parameters of the Stokes and pump photoassociating pulses providing optimal population transfer shown in 
Fig.\ref{resonances1}. We use $q=10$, $\gamma = 10^8$ s$^{-1}$, and $\mu _{2b}=\mu _{21}=0.1$ D 
(1 D=$10^{-18}$ esu cm = 0.3934 $ea_0$). Rabi frequencies are modeled by Gaussians 
$\Omega_{S,p}=\Omega_{S,p}^{0}\exp{(-(t-t_{0}\pm \tau_{S,p}))/T_{S,p}^{2}}$, where $\pm$ refer to the Stokes and pump pulse, respectively.}
\begin{tabular}{c c c c c c c c c c}
\hline
 {\rm Reso-}& $\delta_{\epsilon}$  & $\Gamma$ & $\Omega^0_{S}$    
            & $I_{S}$    & $I_{p}$    & $T_{S}$  & $T_{p}$   & $\tau_{S}$ & $\tau_{p}$ \\
 {\rm nance}& $\mu$K                  & $\mu$K   & $10^{8}$ s$^{-1}$ 
            & W/cm$^{2}$ & W/cm$^{2}$ & $\mu$s   & $\mu$s    & $\mu$s   & $\mu$s     \\
\hline
&&&&&&&&&\\
{\rm None}   & 10      & ---  &  0.72 & 62  & $4\times 10^5$ & 1.5 & 3 & 0.75 & 1.0 \\
{\rm Broad}  & 10      & 1000 &  0.74 & 65 & 4000  & 1.4   & 3.4 & 0.65 & 1.0 \\
{\rm Narrow} & 100     & 1    &  2.24 & 600 & 400  & 0.157 & 0.3 & 0.1  & 0.207 \\
&&&&&&&&&\\
\hline
\end{tabular}
\label{table:res-par}
\end{table}

When comparing the results for a broad resonance to the unperturbed continuum ({\it i.e.}, far from the resonance), we find that 
the source term $S$ is enhanced by the factor $g(q,\epsilon_{0})$ (see Eq. (\ref{eq:S-wide}) in \ref{appendixA}):
\begin{equation}
  g(q,\epsilon_0)= \frac{q+\frac{2}{\Gamma}(\epsilon_0-\epsilon_{F})}{\sqrt{1+\frac{4}{\Gamma^2}(\epsilon_0-\epsilon_{F})^{2}}} .
\end{equation}
This factor has a maximum at $2(\epsilon_{0}-\epsilon_{F})/\Gamma=1/q$, with the corresponding maximum value 
$\sqrt{1+q^2}\approx q$ for $q\gg 1$: hence, the source amplitude is enhanced $q$ times.
In this limit, all populated continuum states experience the same transition dipole matrix element enhancement factor to 
the state $\ket{2}$, so that the system essentially reduces to the case of a flat continuum with an uniformly enhanced transition 
dipole matrix element. One thus expects that in this limit, the adiabatic passage should be efficient, requiring less pump 
laser intensity when compared to the unperturbed ({\it i.e.} without resonance) scattering continuum. This is clearly demonstrated 
in Fig.~\ref{resonances1} (left column, dashed lines): to reach the same $\sim 97$\% transfer efficiency achieved with the 
broad resonance, a very large pump laser intensity is required if there is no resonance in the continuum 
(Fig.~\ref{resonances1} a), while the Stoke laser intensity is basically the same. So, the comparison of Rabi frequencies for the 
broad resonance and no resonance cases shows that, to achieve the same transfer efficiencies, the required peak pump pulse 
intensity is about $100$ times larger without resonance. Condering the intensity used in this particular example, this would lead to intensities in the range of $5\times 10^5$ W/cm$^2$, making STIRAP from the continuum technically impossible to achieve without a resonance. This is consistent with the analysis of photoassociative adiabatic passage from an unstructured continuum \cite{Ye-photoass}, and the above prediction that in the presence of a wide resonance the required 
pump laser intensity is reduced by a factor of $\sim 1/q^{2}$.

\begin{figure}
\center{
\includegraphics[width=8.cm]{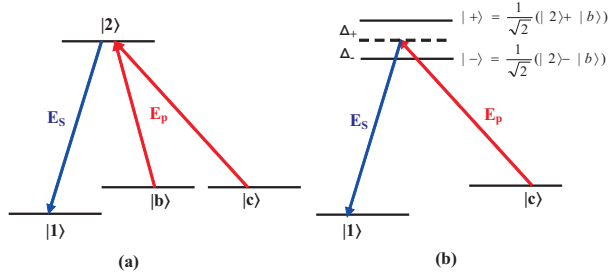}
\caption{\label{fig:interference} Illustration of the reduction of STIRAP transfer efficiency due to destructive quantum interference for a narrow resonance: (a) a simplified level scheme where the scattering continuum is modeled by a single state $\ket{c}$ and the interaction between the continuum and the Feshbach state $\ket{b}$ is neglected; (b) an equivalent scheme, where the strong coupling between the Feshbach state $\ket{b}$ and the excited state $\ket{c}$ by the pump field forms "dressed" states $\ket{\pm}$. We habe the Rabi frequency $\Omega_{2c}=\Omega_{+c}^{2}/\Delta_{+}+\Omega_{-c}^{2}/\Delta_{-}=0$, since $\Omega_{+c}=\Omega_{-c}$ and $\Delta_{+}=-\Delta_{-}$.}
}
\end{figure}

Results of adiabatic passage in a narrow resonance limit are shown in Fig.~\ref{resonances1} (right column). 
We considered a typical value of $\Gamma=1$ $\mu$K for a narrow resonance (see examples in Appendix A.2) and the ensemble energy bandwidth $\delta_{\epsilon}=100$ 
 $\mu$K. Again, we give the parameters providing the optimal transfer in Table~\ref{table:res-par}. In this limit, the transfer efficiency is lower: in the specific 
case analyzed here, it does not exceed 47\%. The reason for this lower efficiency is destructive quantum interference which 
leads to electromagnetically induced transparency \cite{EIT} in the transition from the continuum to the excited state. 
It can be explained using the following argument (see Fig. \ref{fig:interference}). The limit of a narrow Feshbach resonance 
corresponds to a weak coupling between the bound Feshbach state and the scattering continuum, and thus can be neglected in 
this simplified explanation. The system then can be viewed as consisting of bound and continuum states $\ket{b}$ and $\ket{c}$ 
having the same energy, which are coupled by the pump field to a molecular state $\ket{2}$, itself coupled to the state $\ket{1}$ 
by the Stokes field. Assuming that initially all the population is in the state $\ket{c}$, due to the small interaction strength 
between $\ket{b}$ and $\ket{c}$,  we can eliminate the state $\ket{b}$, taking into account its coupling to $\ket{2}$ by the pump 
laser as the formation of ``dressed" states $\ket{\pm}=(\ket{2}\pm \ket{b})/\sqrt{2}$. If the dipole matrix element 
of the $\ket{b}\rightarrow \ket{2}$ transition is much larger than that of the $\ket{c}\rightarrow \ket{2}$ transition, the 
detuning of the ``dressed" states $|\Delta_{\pm}|=\Omega_{p}^{2b}\gg \Omega_{p}^{2c},\;\Omega_{S}$. As a result, the one-photon 
coupling of $\ket{c}$ to the excited state, as well as two-photon coupling to $\ket{1}$ vanishes, preventing the adiabatic 
transfer. This mechanism is similar to the Fano interference effect, the difference is that the continuum is initially populated. 
One can therefore view it as an inverse Fano effect. The effective dipole matrix element of the $\ket{c}\rightarrow \ket{2}$ 
transition is $\mu_{2c}\sim \mu_{2b}/q\sqrt{\xi}$. In the case we analyzed, $q=10$, $\xi=\Gamma/\sqrt{2}\delta_{\epsilon}=0.007$, and $\mu_{2c}\approx \mu_{2b}$, which gives $\sim 50\%$ transfer efficiency. 

The transfer efficiency increases if the Feshbach state is far detuned from the populated continuum. 
Our calculations show that for a Feshbach state detuning $\gg |\Omega_{2b}|^{2}/\gamma$, the transfer efficiency reaches 
$70\%$ using the laser pulse parameters in Table~\ref{table:res-par}. We note that the smaller intensity of the pump pulse 
used for the narrow resonance, as compared to the broad resonance, is due to the fact that we used the same $q=10$ and assumed 
$\mu_{2b}=0.1$ D for both resonances. From the definition of $q$, it means that the continuum-bound dipole matrix element 
$\mu_{2\epsilon}$ is higher in the narrow than in the broad resonance case we considered. This explains the smaller resulting 
pump pulse intensity. The overall conclusion for a narrow resonance is that, as opposed to a broad resonance, the presence of the 
Feshbach resonance prevents one from realizing high transfer efficiencies. It should be noted, however, that the destructive 
quantum interference effect is based on negligible interaction between the Feshbach and continuum states during the transfer 
time, since $\tau_{\mathrm{tr}} < \delta_{\epsilon}^{-1} \ll \Gamma^{-1}$. This argument shows that 
already for $\Gamma \ge \delta_{\epsilon}$, there is enough interaction to neutralize the effect of destructive interference. 
Therefore, we expect that the broad resonance limit can be extended down to $\Gamma \sim \delta_{\epsilon}$, making it applicable to a wide variety of atomic species.

\section{Applications to the conversion of an entire atomic ensemble into a ground rovibrational molecule gas}

\begin{figure}[h]
\begin{center}
\includegraphics[width=\textwidth,clip]{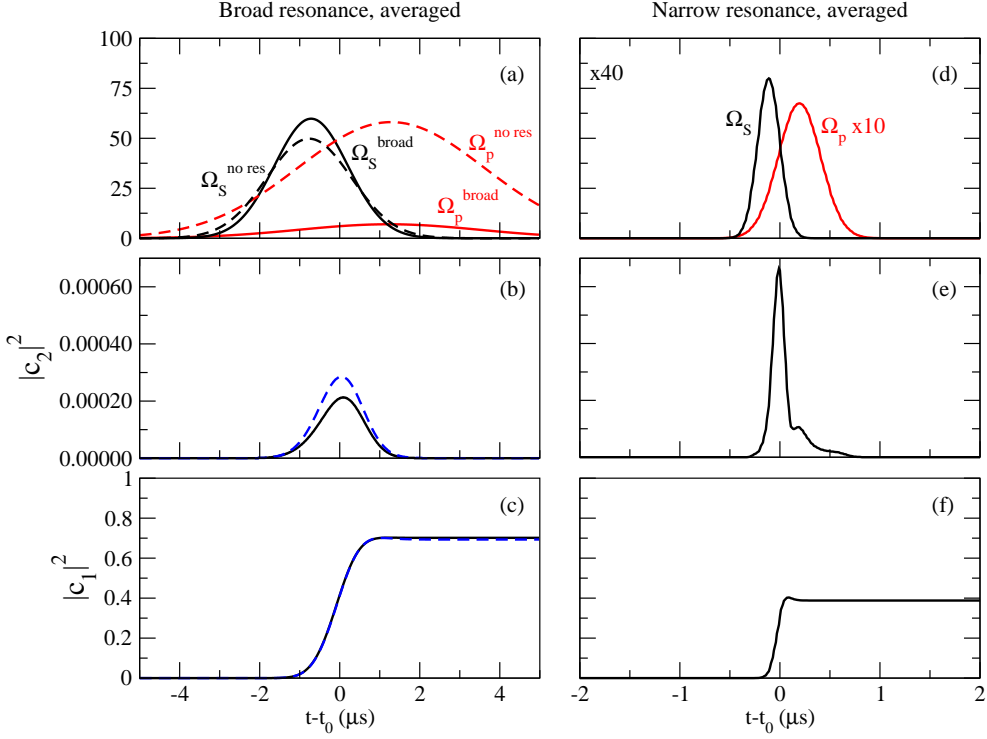}
\end{center}
\caption{Same as Fig.~\ref{resonances1}, but for the energy averaged transfer. The parameters are listed in Table~\ref{table:res-par-avg}.}
\label{resonances2}
\end{figure}

The results of Fig.~\ref{resonances1} are for a pair of atoms having a specific mean collision energy 
$\epsilon_{0}=\hbar(\omega_{S}-\omega_{p})$. Such situation could be realized in very tight traps, {\it e.g.}, 
in tight optical lattices. For a system with a wider energy distribution, one would like to find an ensemble averaged 
transfer efficiency, and thus one needs to calculate the transfer probability $P(\epsilon_{0})=\left|c_{1}\right|^{2}$ for all $\epsilon_{0}$ within the thermal spread of energies, and perform the averaging as
\begin{equation}
P_{\rm avg}=\frac{2}{\sqrt{\pi}(k_{B}T)^{3/2}}\int_{0}^{\infty}e^{-\epsilon_{0}/k_{B}T}\sqrt{\epsilon_{0}}P(\epsilon_{0})d\epsilon_{0},
\end{equation}
where we assume a Maxwell-Boltzmann energy distribution, the pump laser resonant with the center of the 
distribution at $\langle \epsilon \rangle=3/2k_{B}T$, and set the bandwidth of the distribution at 
$\delta_{\epsilon}=\sqrt{\langle (\Delta \epsilon)^{2} \rangle}=\sqrt{3/2}k_{B}T$. The results are shown in 
 Fig.\ref{resonances2}. In this case, while the maximal transfer efficiency in the broad resonance case is $\sim 70\%$, it can be 
achieved with lower laser intensities than in the case of a pair of atoms of Fig. \ref{resonances1}.

\begin{table}
\centering
\caption{Parameters of the Stokes and pump photoassociating pulses providing optimal population transfer shown in Fig.\ref{resonances2} for averaging over a Maxwell-Boltzmann distribution of energies. We use $q=10$, $\gamma = 10^8$ s$^{-1}$, and $\mu _{2b}=\mu _{21}=0.1$ D (1 D=$10^{-18}$ esu cm = 0.3934 $ea_0$).}
\begin{tabular}{c c c c c c c c c c}
\hline
 {\rm Reso-}& $\delta_{\epsilon}$  & $\Gamma$ & $\Omega^0_{S}$    
            & $I_{S}$    & $I_{p}$    & $T_{S}$  & $T_{p}$   & $\tau_{S}$ & $\tau_{p}$ \\
 {\rm nance}& $\mu$K                  & $\mu$K   & $10^{8}$ s$^{-1}$ 
            & W/cm$^{2}$ & W/cm$^{2}$ & $\mu$s   & $\mu$s    & $\mu$s   & $\mu$s     \\
\hline
&&&&&&&&&\\
{\rm None}   & 10      & ---  &  0.50 & 30  & $1.7\times 10^5$ & 1.5 & 3.3 & 0.75 & 1.3 \\
{\rm Broad}  & 10      & 1000 &  0.60 & 40 & 2500  & 1.3   & 3.2 & 0.7 & 1.25 \\
{\rm Narrow} & 100     & 1    &  2.24 & 600 & 400  & 0.157 & 0.3 & 0.1  & 0.207 \\
&&&&&&&&&\\
\hline
\end{tabular}
\label{table:res-par-avg}
\end{table}

Given the adiabatic photoassociation probability $P(\epsilon)$ for two colliding atoms with relative energy $\epsilon$, we can calculate the number of atoms photoassociated during 
the time overlap $\tau_{\mathrm{tr}}$ of the Stokes and pump pulses. During this time, the atom with the energy $\epsilon=\mu v^{2}/2$, where 
$\mu$ is the reduced mass, will collide with atoms in the 
volume $\pi b^{2}v \tau_{\mathrm{tr}}$, where $\pi b^{2}$ is the collision cross-section. The impact parameter for the collision corresponding 
to a partial wave with angular momentum $J$ is $b=(J+1/2)\hbar/p=(J+1/2)\hbar/\sqrt{2\mu\epsilon}$. The number of collisions that 
atoms with a relative energy in the interval $\left(\epsilon,\epsilon+d\epsilon\right)$ will experience during the transfer 
time is therefore $N(\epsilon)d\epsilon=\pi b^{2} v \tau_{\mathrm{tr}}\rho(\epsilon)d\epsilon$, 
where $\rho(\epsilon)=2\rho \exp{(-\epsilon/k_{B}T)}\sqrt{\epsilon}/\sqrt{\pi}(k_{B}T)^{3/2}$ is the spectral density of the 
atoms ($\rho$ is the density of the sample). Finally, $J=0$ for ultracold $s$-wave collisions, and the fraction of atoms in the 
energy interval $\left(\epsilon,\epsilon+d\epsilon\right)$ photoassociated by the two pulses is $f(\epsilon)=P(\epsilon)N(\epsilon)$, or
\begin{equation}
f(\epsilon)=\frac{\sqrt{2\pi} \hbar^{2}}{4(\mu k_{B}T)^{3/2}}\tau_{\mathrm{tr}}\rho P(\epsilon)\exp{(-\epsilon/k_{B}T)}.
\end{equation} 
The total fraction of atoms photoassociated by a pair of pulses is 
$f=\int_{0}^{\infty} d\epsilon\; f(\epsilon) \approx \langle P_{\rm avg} \rangle \rho \sqrt{2\pi} \tau_{\mathrm{tr}}\hbar^{2}/4\mu^{3/2}\sqrt{k_{B}T}$, 
where we assumed that $P(\epsilon)$ does not significantly vary within the ensemble, and approximated it by the averaged value. 
Considering as an example $^{6}$Li atoms at T=100 $\mu$K with an atomic density $\rho=10^{12}$ cm$^{-3}$, an overlap 
time $\tau_{\mathrm{tr}}\sim 1$ $\mu$s, and assuming $P_{\rm avg}=0.7$, the fraction of atoms photoassociated by the Stokes and pump pulses is $f \sim 2.5\times 10^{-4}$: for heavier atoms $f \sim 10^{-6}-10^{-5}$. It will therefore require $\sim 10^{4}-10^{6}$ pairs of pulses to convert an entire atomic ensemble into deeply bound molecules.

Since only a small fraction of atoms can be transferred to $\ket{1}$ by a pair of STIRAP pulses, a train of pulse 
pairs can be applied to photoassociate the entire atomic ensemble. To prevent excitation of molecules in $\ket{1}$ by subsequent 
pulses, they have to be removed before the next pair of pulses is applied. This could be realized by applying, after each pair 
of Stokes and pump pulses, a relatively long pulse resonant to a transition from $\ket{1}$ to some other vibrational level in 
the excited electronic potential which decays spontaneously to a deep vibrational state in the ground electronic potential. 
This long pulse would optically pump molecules out of the state $\ket{1}$ to deeper vibrational states in the ground electronic 
potential. It therefore has to be longer than the spontaneous decay time of the excited state. Care has to be taken that the 
excited state does not decay back into the scattering continuum. This would empty the $\ket{1}$ state and deposit molecules into ground potential vibrational states according to Franck-Condon factors before 
the next pair of pulses arrives. Finally, after all atoms have been converted into molecules the recently demonstrated optical pumping for molecules method \cite{opt-pump} can be applied, which would transfer molecules from all populated vibrational states into the ground level $v=0$. 

The optimal strategy is to actually choose an excited state that decays mostly to the $v=0$ level. This would allow one to avoid storing molecules in unstable vibrational states and using the optical pumping method. If such a state cannot be directly reached from $\ket{1}$, a four-photon STIRAP transfer can be applied \cite{Our-PRA}, which provides efficient transfer to deeply bound molecular 
states. It allows one to choose the final state $\ket{1}$, from which the excited state decaying predominantly to $v=0$ can be easily reached. In this case rotational selectivity can also
be preserved, since only $v=0,J=0$ and $v=0,J=2$ states will be populated. 

The total time required to photoassociate the whole atomic ensemble and transfer it to the $v=0$ level can be estimated as 
follows. As the numerical results show, adiabatic passage requires $\sim 5$ $\mu$s, the follow-up pulse emptying state 
$\ket{1}$ can have a $\sim 100$ ns duration, if the excited state lifetime is tens of ns, resulting in the whole sequence $\sim 6$ $\mu$s. Then the train of $10^{4}-10^{6}$ pulse pairs will take $\sim 0.1-10$ s. The final step, optical pumping to the $v=0$ level, requires $\sim$ hundred $\mu$s, so the overall formation time is $\sim 0.1-10$ s. Given 
an illuminated volume $\sim 1$ mm$^{3}$ and an atomic density $\rho \sim 10^{12}$ cm$^{-3}$ the resulting production rate is expected to be $10^{8}-10^{10}$ molecules/s. This compares well with the recent experiment on STIRAP production of ground state KRb molecules starting from the Feshbach state, where the entire cycle including creation of Feshbach molecules takes $\sim 10$ s \cite{STIRAP-Fesh}.

\section{Conclusion}

Combining photoassociation and coherent optical transfer to molecular ground vibrational states can allow one to convert an 
entire atomic ensemble into deeply bound molecules, and to produce a 
high phase-space density ultracold molecular gas. We have analyzed photoassociative adiabatic passage in a thermal ultracold 
atomic gas near a Feshbach resonance. The presence of a bound state imbedded in and resonant with scattering continuum states strongly enhances the continuum-bound transition dipole matrix element to an excited electronic state, thus requiring less laser intensity for efficient transfer. In the limit of a wide resonance when compared to the thermal spread of collision 
energies, the dipole matrix element is enhanced by the Fano parameter $q$. Choosing a tightly bound excited vibrational state, $q$ can be made much larger than unity, resulting in the intensity of the pump pulse required for efficient adiabatic passage to be $\sim 1/q^{2}$ times smaller than in the absence of the resonance. We modeled the adiabatic passage using typical parameters of alkali dimers and found intensities and durations of STIRAP pulses providing optimal transfer. Intensities of the pump pulse, coupling the continuum to an excited state, were found to be a few kW/cm$^{2}$, which is $\sim 100$ times smaller than without resonance. Optimal pulse durations are several $\mu$s, resulting 
in energies per pulse $\sim 10$ $\mu$J for a focus area of $1$ mm$^{2}$.  

If the Feshbach resonance is narrow compared to the thermal energy spread of colliding atoms, adiabatic passage is hindered by destructive quantum interference. The reason is that electromagnetically induced transparency significantly reduces the transition dipole matrix element from the scattering continuum to an excited state in the presence of the bound Feshbach state. In the 
narrow resonance limit, photoassociative adiabatic passage is therefore more efficient if the resonance is far-detuned.

Due to low atomic collision rates at ultracold temperatures, only a small fraction of atoms can be converted into molecules by a pair of photoassociative pulses. To convert an entire atomic ensemble, a train of pulse pairs can be applied. We estimate that $10^{4}-10^{6}$ pulse pairs will associate an atomic gas of alkali dimers with a density $10^{12}$ cm$^{-3}$ in an illuminated volume of $1$ mm$^{3}$ in $0.1-10$ s, resulting in extremely high production rates of $10^{8}-10^{10}$ molecules/s. 
High transfer efficiencies combined with low intensities of adiabatic photoassociative pulses also make the broad resonance limit attractive for quantum computation. For example, a scheme proposed in \cite{Ostrovskaya} can be realized, where qubit states are encoded into a scattering and a bound molecular states of polar molecules. To perform one and two-qubit operations, this scheme requires a high degree of control over the system, which our model readily offers.

Finally, marrying FOPA and STIRAP is a very promising avenue to produce large amounts of molecules, for a variety of molecular 
species. In fact, although we described here examples based on magnetically induced Feshbach resonances, such resonances are 
extremely common, and can be induced by several interactions, such as external electric fields or optical fields. Even in the 
absence of hyperfine interactions, other interactions can provide the necessary coupling, such as in the case of the magnetic 
dipole-dipole interaction in $^{52}$Cr \cite{pfau,pavlovic}.

\section*{Acknowledgments}

This research was partially founded by the National Science Foundation, Army Research Office, and the U.S. Department of Energy,
Office of Basic Energy Sciences.

\appendix
\section{Adiabatic passage in the limits of broad and narrow Feshbach resonances}
\label{appendixA}

In this appendix, we discuss Eqs.(\ref{eq:c1-new}) and (\ref{eq:c2-new}) for various relative widths of the Feshbach 
resonance $\Gamma$ with respect to the thermal energy spread $\delta_{\epsilon}$ of the colliding atoms. We 
first describe the case of a broad resonance, {\it i.e.}, when the width of the Feshbach resonance greatly exceeds the thermal energy spread ($\Gamma\gg\delta_{\epsilon}$), and second consider the opposite situation of a narrow resonance ($\Gamma\ll\delta_{\epsilon}$). Finally, we briefly present the case where there is no resonance.

\subsection{Limit of a broad Feshbach resonance $\Gamma\gg \delta_{\epsilon}$}

The typical thermal energy spread for colliding atoms in photoassociation experiments with non-degenerate gases is 
$\delta_{\epsilon} \sim 10-100$ $\mu$K. The broad resonance case occurs for resonances having a width of several Gauss ($\sim 1$ mK), 
for which we have $\Gamma/\delta_{\epsilon}\sim 10 -100$. A wide variety of systems exhibit broad resonances. For instance, they can be found in collision of $^{6}$Li atoms at 834 G for the $|f=1/2,m_{f}=1/2\rangle$ channel ($\Gamma = 302$ G= 40 mK) and in $^{7}$Li at 736 G for the $|f=1,m_{f}=1\rangle$ channel ($\Gamma = 145$ G = 19 mK).
We note here that these values of $\Gamma$ are slightly different than the ``magnetic" width $\Delta B$ usually given and based on the modelling of the scattering length.

The source function can be readily calculated from Eq.(\ref{eq:S}) by noticing that the Rabi frequency term can be 
set at $\epsilon=\epsilon_{0}$ corresponding to the maximum of the Gaussian function in the integrand. Using the function 
$g(q,\epsilon)$ defined in Eq.(\ref{eq:g}), the result takes the form 
\begin{eqnarray}
\label{eq:S-wide}
S_{w}&=&S_{0}\sqrt{2\pi}\delta_{\epsilon} g(q, \epsilon_{0})\mathrm{sgn}(\epsilon_{0}-\epsilon_{F}) e^{-(t-t_{0})^{2}\delta^{2}_{\epsilon}/2\hbar^{2}-i(\epsilon_{0}/\hbar-(\omega_{S}-\omega_{p}))t} \nonumber \\
&=&S_{\mathrm{no-res}}g(q, \epsilon_{0})\mathrm{sgn}(\epsilon_{0}-\epsilon_{F}),
\end{eqnarray}
where $S_{\mathrm{no-res}}$ is the source function without a resonance given below in Eq.(\ref{eq:sorce-nr}). 
Strictly speaking, this expression is valid for $|\epsilon_{F}-\epsilon_{0}|\ge \delta_{\epsilon}$, but since 
$\Gamma \gg \delta_{\epsilon}$ Eq.(\ref{eq:S-wide}) is a good approximation for a wide range of detunings $\epsilon_{F}-\epsilon_{0}$.

The back-stimulation term (\ref{eq:back-stimulation}) can be further simplified in the limit of a broad resonance. 
In this case, both $c_{2}(t)$ and ${\cal E}_{p}(t)$ change on a time scale 
$\sim 1/\delta_{\epsilon}$, {\it i.e.}, slowly compared to the decay time $\sim \hbar/\Gamma$ of the exponent. Therefore, we can rewrite (\ref{eq:back-stimulation}) as:
\begin{equation}
\left|\frac{\vec{\mu}_{2\epsilon}\hat{\vec{e}}_{p}}{\hbar}\right|^{2}\pi \hbar \left[1+\frac{(q-i)^{2}}{1+2i(\epsilon_{F}-\hbar(\omega_{S}-\omega_{p}))/\Gamma}\right]c_{2}(t){\cal E}^{2}_{p}(t).
\end{equation}
The system (\ref{eq:c1-new})-(\ref{eq:c2-new}) in the case of a broad resonance becomes:
\begin{eqnarray}
\label{eq:c1-new1}
i \frac{\partial c_{1}}{\partial t} & = & -\Omega_{S}c_{2},\\
\label{eq:c2-new1}
i\frac{\partial c_{2}}{\partial t} & = & -\Omega_{S}c_{1} - S_{w}+(\delta -i\gamma)c_{2} \nonumber\\
 &  & -i\pi \hbar|\Omega_{\mathrm{no-res}}(t)|^{2}\left[1+\frac{(q-i)^{2}}{1+2i(\epsilon_{F}-\hbar(\omega_{S}-\omega_{p}))/\Gamma}\right]c_{2},
\end{eqnarray} 
where $\Omega_{\mathrm{no-res}}=\vec{\mu}_{2\epsilon}\hat{\vec{e}}_{p}{\cal E}_{p}/\hbar$ is the continuum-bound Rabi frequency in 
the absence of resonance. We also added a spontaneous decay term $\gamma c_{2}$, assuming that the excited molecules 
dissociate into high energy continuum states and the resulting atoms leave a trap.
From Eq.(\ref{eq:S-wide}), one can see that in a broad resonance case, the 
source amplitude is enhanced by the factor 
$g(q,\epsilon_{0})=(q+2(\epsilon_{0}-\epsilon_{F})/\Gamma)/\sqrt{1+4(\epsilon_{0}-\epsilon_{F})^{2}/\Gamma^{2}}$ when compared 
to the unperturbed continuum case. This factor has a maximum at $2(\epsilon_{0}-\epsilon_{F})/\Gamma=1/q$, with the corresponding maximum value $g_{\rm max}\sim q$ for $q\gg 1$.

\subsection{Limit of a narrow Feshbach resonance $\Gamma \ll \delta_{\epsilon}$}

This situation occurs when the width of the resonance is of the order of a few micro-Gauss or less. 
Examples of narrow resonances include $^{6}$Li$^{23}$Na at 746 G for the 
$|f_{1}=1/2,m_{f1}=1/2\rangle |f_{2}=1,m_{f2}=1\rangle$ channel ($\Gamma=7.8$ mG = 1 $\mu$K) \cite{LiNa}, or $^{6}$Li$^{87}$Rb at 882 G for the $|f_{1}=1/2,m_{f1}=1/2\rangle |f_{2}=1,m_{f2}=1\rangle$ channel (p-wave, $\Gamma =10$ mG = 1.3 $\mu$K).

We note that the source term expressed in Eq.(\ref{eq:S}) can be rewritten in a time representation: 
\begin{eqnarray}
\label{eq:source}
S&=&S_{0}\sqrt{2\pi}\delta_{\epsilon}e^{-i(\epsilon_{0}/\hbar-(\omega_{S}-\omega_{p}))t} \nonumber \\
 &\times &  \left[ e^{-(\tau -\tau_{0})^{2}} + \xi e^{2iD-D^{2}}\int_{-\infty}^{\infty}e^{-(\tau'-iD)^{2}}(I_{1}(\xi|\tau-\tau_{0}-\tau'|) \right. \nonumber \\ 
&& -L_{-1}(\xi|\tau-\tau_{0}-\tau'|)-iq(I_{0}(\xi|\tau-\tau_{0}-\tau'|) \nonumber \\
&& \left.-L_{0}(\xi|\tau-\tau_{0}-\tau'|))\mathrm{sgn}(\tau-\tau_{0}-\tau'))d\tau'\right] , 
\end{eqnarray}
where we introduced the dimensionless variables $\tau=t\delta_{\epsilon}/\sqrt{2}\hbar$, $D=(\epsilon_{F}-\epsilon_{0})/\sqrt{2}\delta_{\epsilon}$, 
$\xi=\Gamma/\sqrt{2}\delta_{\epsilon}$; $I_{0,1}$ and $L_{0,-1}$ are modified Bessel and Struve functions. 
One can see from this expression that the source function is a sum 
of the pure source function of the unperturbed continuum, given by the first term in square brackets, and of the admixed 
bound state, given by the integral. The coefficient $\xi=\Gamma/\sqrt{2}\delta_{\epsilon}$, which is the ratio of the Feshbach resonance width to the width of the thermal energy spread, gives the ratio of contributions from the bound state and the unperturbed continuum, respectively. 

It is then easier to notice that in the limit of a narrow resonance, the Gaussian function in the integrand of Eq.(\ref{eq:source}) is much narrower than the Bessel and Struve functions, which change on the time scale $\sim 1/\xi$. Therefore the source term can be aproximated as:
\begin{eqnarray}
\label{eq:S-narrow-res}
S_{n}&=&S_{0}\sqrt{2\pi}\delta_{\epsilon}e^{-i(\epsilon_{0}/\hbar-(\omega_{S}-\omega_{p}))t}
[e^{-(\tau -\tau_{0})^{2}}  \nonumber \\
&&+\xi \sqrt{\pi}e^{2iD-D^{2}}(I_{1}(\xi|\tau -\tau_{0}|)-L_{-1}(\xi|\tau -\tau_{0}|) \nonumber \\
&&-iq(I_{0}(\xi|\tau -\tau_{0}|)-L_{0}(\xi|\tau -\tau_{0}|))\mathrm{sgn}(\tau -\tau_{0}))]. 
\end{eqnarray}

Since $\xi \ll 1$, the real part of the source function is given by the first term in the square brackets, which is a pure continuum source function, while the imaginary part is due to the admixed bound state and its magnitude depends on the product $\xi q$. Using asymptotic expansions of modified Bessel and Struve functions $I_{0}(x)-L_{0}(x)\rightarrow -2/\pi x$, $I_{1}(x)-L_{-1}(x)\rightarrow -2/\pi x^{2}$, it is seen from Eq.(\ref{eq:S-narrow-res}) that the contribution to the source function from the bound state decays on the time scale $|\tau -\tau_{0}|\sim 1/\xi$, while the contribution from the unperturbed continuum decays on the time scale $|\tau -\tau_{0}|\sim 1 \ll 1/\xi$. 

In the limit of a narrow resonance the system (\ref{eq:c1-new})-(\ref{eq:c2-new}) becomes:

\begin{eqnarray}
\label{eq:c1-new2}
i \frac{\partial c_{1}}{\partial t} & = & -\Omega_{S}c_{2},\\
\label{eq:c2-new2}
i\frac{\partial c_{2}}{\partial t} & = & -\Omega_{S}c_{1}-S_{n}+ (\delta -i\gamma)c_{2} \nonumber\\
&&-i\left|\frac{\vec{\mu}_{2\epsilon}\hat{\vec{e}}_{p}}{\hbar}\right|^{2}\left[\pi \hbar{\cal E}_{p}^{2}c_{2}+\frac{\pi\Gamma}{2}(q-i)^{2}{\cal E}_{p}(t)\right. \nonumber \\
 & & \left.\times\int_{0}^{t} dt'\; c_{2}(t'){\cal E}_{p}(t')e^{\Gamma(t'-t)/2\hbar
+i(\epsilon_{F}/\hbar-(\omega_{S}-\omega_{p}))(t'-t)}\right] .
\end{eqnarray}

\subsection{Continuum without resonance}

Finally, let us consider the case of a continuum without resonance. In this case the continuum-bound Rabi frequency Eq.(\ref{eq:Rabi-fr}) is:
\begin{equation}
\Omega_{\epsilon}=\Omega_{\mathrm{no-res}}=\vec{\mu}_{2\epsilon}\cdot\hat{\vec{e}}_{p}\;{\cal E}_{p}/\hbar ,
\end{equation}
and the source function is
\begin{equation}
\label{eq:sorce-nr}
 S_{\mathrm{no-res}}=S_{0}\sqrt{2\pi}\delta_{\epsilon}\e^{-(t-t_{0})^{2}\delta_{\epsilon}^{2}/2\hbar^{2}-i(\epsilon_{0}/\hbar -(\omega_{S}-\omega_{p}))t} .
\end{equation}

The back-stimulation term (\ref{eq:back-stimulation}) reduces to
\begin{equation}
\left|\vec{\mu}_{2\epsilon}\cdot\hat{\vec{e}}_{p}/\hbar\right|^{2}\pi \hbar {\cal E}_{p}^{2}c_{2}=\pi \hbar \left|\Omega_{\mathrm{no-res}}(t)\right|^{2}c_{2} ,
\end{equation}
and the system (\ref{eq:c1-new})-(\ref{eq:c2-new}) takes the simple form:
\begin{eqnarray}
\label{eq:c1-no-res}
i \frac{\partial c_{1}}{\partial t} & = & -\Omega_{S}c_{2},\\
\label{eq:c2-no-res}
i\frac{\partial c_{2}}{\partial t} & = & -\Omega_{S}c_{1} + (\delta -i\gamma)c_{2}
-i\pi \hbar|\Omega_{\mathrm{no-res}}(t)|^{2}c_{2}-S_{\mathrm{no-res}}. 
\end{eqnarray}

\end{document}